\documentstyle[sprocl]{article}
\input epsf

\def\be{\begin{equation}}
\def\ee{\end{equation}}
\def\bea{\begin{eqnarray}}
\def\eea{\end{eqnarray}}

\newenvironment{indention}[1]{\par
\addtolength{\leftskip}{#1}
\begingroup}{\endgroup\par}

\begin{document}

\title{DUAL HIGGS THEORY FOR COLOR CONFINEMENT}

\author{H. ICHIE}

\address{Department of Physics, Tokyo Institute of Technology \\
Meguro, Tokyo 152-8551, Japan
\\E-mail: ichie@th.phys.titech.ac.jp} 

\author{H. SUGANUMA}

\address{Research Center for Nuclear Physics (RCNP), Osaka University \\
Ibaraki, Osaka 567-0047, Japan
}


\maketitle
\abstracts{ 
 We study theoretical bases of the dual Higgs theory for
 confinement physics in QCD in terms of monopoles and the gluon
 configuration in the maximally abelian (MA) gauge.
 Abelian dominance for the confinement force can be
 analytically proved by regarding the off-diagonal
 angle variable as a random variable in the lattice formalism.
 In the long-distance scale, the contribution of off-diagonal
 gluons to the Wilson loop cancels each other and exhibits a perimeter
 law behavior, which leads to  exact abelian dominance
 on the string tension if the finite size effect of the Wilson loop is removed.
 We investigate the appearance of the monopole in the QCD vacuum,
 considering the role of off-diagonal gluons.
 The monopole carries a large fluctuation of the gluon field
 and provides a large abelian action in abelian projected QCD.
 Due to the partial cancellation between the abelian part and the off-diagonal 
 part of the QCD action, the monopole can appear in QCD without large cost of 
 the QCD action. 
 The off-diagonal gluon is necessary
 for existence of the monopole at the short-distance scale.
 We study monopole condensation, which is 
 the requirement 
 of the dual Higgs 
 theory, by comparing the QCD vacuum with the monopole-current system.
 We find that ``entropy'' of monopole-current dominates than  its  ``energy'', 
 and the monopole seems to be condensed at the infrared scale in the QCD vacuum.
}

\section{Dual Superconductor Picture for Confinement in QCD}

The strong interaction is subjected to Quantum 
Chromodynamics (QCD).
Due to the self-interaction of the gauge field,  
the QCD gauge coupling becomes very strong in 
the low-energy region, while it is weak in the high-energy region.
Accordingly, QCD phenomena are divided into two theoretical categories:
the strong coupling leads to complicated nonperturbative phenomena such as
color confinement and chiral symmetry breaking, while
high-energy phenomena are understood by the perturbative QCD.
Since there is no well-established analytical method for
nonperturbative phenomena,
one must carry out the Monte Carlo simulation based on  the lattice QCD
or apply the effective models described by the relevant degrees of freedom 
for the low-energy physics. 
As for the chiral dynamics, 
the pion and the sigma meson, 
which are composite modes of quark and anti-quark,
play an important role
for the infrared effective theory  such as 
 the (non-)linear sigma model and the Nambu-Jona-Lasinio model.
On the other hand, confinement is essentially described by the dynamics of 
gluons rather than quarks.

\vspace{-0.01cm}

In 1970's, Nambu 
proposed an interesting idea that quark confinement can be interpreted 
using the dual version of the 
superconductivity\cite{nambu}. 
In the 
superconductor, 
Cooper-pair condensation leads to the Meissner effect, 
and the magnetic flux is 
squeezed like a 
quasi-one-dimensional tube as the Abrikosov vortex.
On the other hand, 
from the Regge trajectory of hadrons and the lattice QCD, 
the confinement force between the color-electric charge is 
brought by one-dimensional squeezing of 
the color-electric flux
in the QCD vacuum. 
Hence, the QCD vacuum can be regarded as the dual version 
of the superconductor.
In this dual-superconductor picture for the QCD vacuum, 
the squeezing of the color-electric flux between quarks 
is realized by the dual Meissner effect
as the result of condensation of color-magnetic monopoles.
However, there are two 
large gaps between QCD and the dual 
superconductor picture. 
\vspace{-0.2cm}
\begin{enumerate}
\item
This picture is based on the abelian gauge theory, 
while QCD is a nonabelian gauge theory. 
\vspace{-0.2cm}
\item
The dual-superconductor scenario requires condensation of magnetic \\
monopoles as the key concept, while QCD does not have such a monopole as 
the elementary degrees of freedom. 
\end{enumerate}
\vspace{-0.2cm}
\indent As the connection between QCD and the dual superconductor scenario, 
't Hooft proposed the concept of the abelian gauge fixing\cite{thooft}, 
the partial gauge
\newlength{\miniwocolumn}
\setlength{\miniwocolumn}{0.5\textwidth}
\begin{minipage}[t]{\miniwocolumn}
 fixing which is defined so as to diagonalize a suitable 
gauge-dependent variable $\Phi[A_\mu(x)]$.
The abelian gauge fixing reduces QCD into an abelian gauge
theory, where the off-diagonal element of the gluon field behaves as a 
charged matter field (See Fig.1).
As a remarkable fact in the abelian gauge, color-magnetic monopoles appear
as topological objects corresponding to 
the nontrivial homotopy group $\Pi_2( {\rm SU(N_c)/U(1)}^{\rm N_c-1}) =
{\bf Z}^{\rm N_c-1}_\infty$.
Here, we assume  abelian
 dominance, 
 which means
that the only abelian gauge fields with monopoles 
would be essential for the description of the
 nonperturbative QCD,
 and neglect the off-diagonal elements, which is called  
\end{minipage}
\begin{minipage}[t]{\miniwocolumn}
\vspace{-0.8cm}
\epsfxsize = 5 cm
\centering \leavevmode
\epsfbox{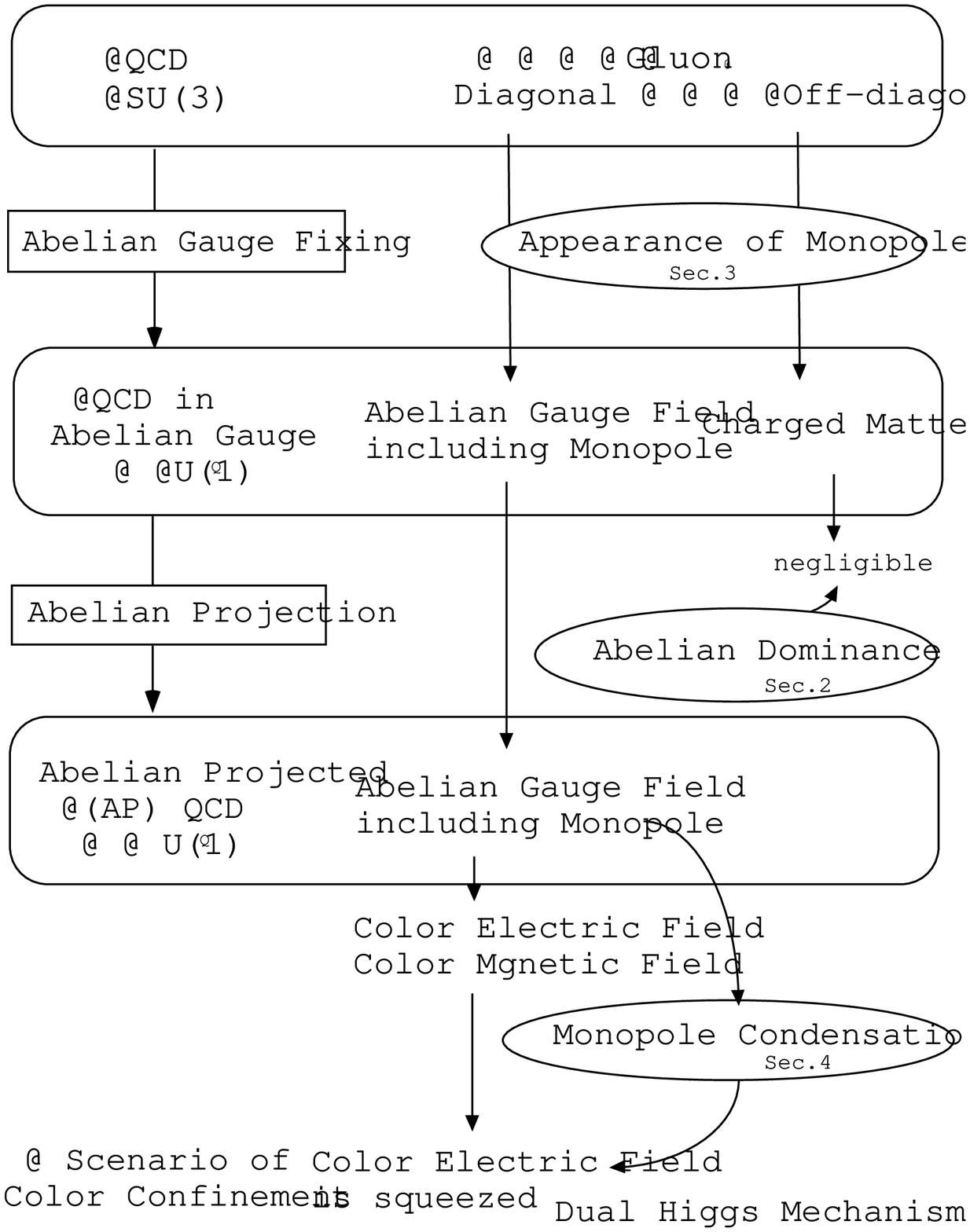}

\hspace{0.2cm}
{\footnotesize {\bf Fig.1} Dual Superconductor Picture \\ in QCD
}
\label{aim}
\vspace{0cm}
\end{minipage}
 abelian 
projection. Thus, by the abelian gauge fixing and the abelian projection, 
QCD is reduced into  abelian projected (AP-)QCD, which is abelian gauge 
theory including monopoles.
If the monopole condenses, 
the scenario of color confinement by the dual Meisser effect would be 
a realistic picture for confinement 
in QCD.
In this paper, using the lattice QCD simulation,
we study the key of the dual Higgs theory: abelian dominance (Sec.2), 
the appearance of monopole (Sec.3) and monopole condensation (Sec.4).

\section{Abelian Dominance in the Maximally Abelian Gauge}

Abelian dominance on the confinement force 
has been investigated using 
the lattice QCD simulation in the maximally abelian (MA)
gauge\cite{giacomo,poly}.
In terms of the link variable $U_\mu(s) \equiv U_\mu^0(s) + i\tau^a 
U_\mu^a(s)$, the MA gauge fixing is defined by maximizing 
$R \equiv \sum_{s,\mu} {\rm tr}( U_\mu(s) \tau_3 U^{\dagger}_\mu(s) \tau_3 ) 
= \sum_{s,\mu}\{ 
(U^0_\mu(s))^2+(U^3_\mu(s))^2-(U^1_\mu(s))^2-(U^2_\mu(s))^2 \}$
through the gauge transformation.
In this gauge, the off-diagonal components, $U_\mu^1$ and $U_\mu^2$, are 
forced to be small,
and therefore the QCD system seems describable by U(1)-like variables 
approximately.
It is to be noted that the MA gauge is a sort of the abelian gauge, because
$\Phi(s) \equiv \sum _{\mu,\pm } U_{\pm \mu}(s) \tau_3 U^{\dagger}_{\pm \mu}(s)$
is diagonalized there. (For the simple notation, we use $U_{-\mu}(s) \equiv 
U^{\dagger}_\mu(s-\mu)$ in this paper).
In this section, we study the origin of
abelian dominance on the confinement force in the QCD vacuum, 
considering the relation to abelian dominance on the link variable.

\vspace{-0.3cm} 

\subsection{Abelian Dominance on  Link Variable: 
Microscopic Abelian Dominance} 

\vspace{-0.1cm} 

\indent\indent In the lattice formalism, the SU(2) link variable 
$U_\mu$(s) is
factorized as \\

\vspace{-0.2cm}
{\small
\noindent $U_\mu(s) 
=\left( {\matrix{
{\rm cos}{\theta_\mu}(s) & -{\rm sin}{\theta_\mu}(s) e^{-i\chi_\mu(s)} \cr
{\rm sin}{\theta_\mu}(s) e^{i\chi_\mu(s)} & {\rm cos}{\theta_\mu}(s)
}} \right)$$\left( \begin{array}{cc} e^{i\theta^3_\mu(s)} & 0 \\
 0 & e^{-i\theta^3_\mu(s)}
 \end{array} \right) 
\equiv M_\mu (s) u_\mu(s), \nonumber
$} \\ 

\noindent where $u_\mu(s)$ and  $M_\mu(s)$ correspond to the diagonal part and the 
off-diagonal part, respectively.
For the residual U(1)$_3$ gauge transformation, $u_\mu(s)$ behaves as the 
abelian gauge field, while $c_\mu(s)$ behaves as the charged matter.
Here, the range of angle variables are taken as $0 \le \theta_{\mu} 
\le \frac{\pi}{2}$, $-\pi \le \theta_{\mu}^3, \chi_\mu  < \pi$.

In order to investigate  abelian dominance on the link variable in the 
MA gauge,
we define ``abelian projection rate'' as $R_{\rm Abel}(s,\mu) \equiv \cos 
\theta_\mu(s)$\cite{suganuma1}. 
For instance, if $\cos \theta_{\mu}  =1$,
the SU(2) link variable is completely abelian as
{\small $U_\mu(s) = 
 \left( {\matrix{
 e^{i\theta^3_\mu}
  &          0
\cr
0                     &    e^{-i\theta^3_\mu} \cr
}} \right),
$} 
while  it is 
off-diagonal
 as 
{\small $U_\mu(s)  =
 \left( {\matrix{
0 & -e^{-i(\theta^3_\mu+\chi_\mu)}  
\cr
e^{i(\theta^3_\mu+\chi_\mu)} & 0\cr
}} \right)  $ }  if  $\cos \theta_{\mu} =0$.
We show in Fig.2 the spatial distribution of the abelian projection rate
$R_{\rm Abel} = \cos \theta$
as an arrow $(\sin \theta, \cos \theta)$.
In the MA gauge, most of all SU(2) link variables become U(1)-like.
We show also in Fig.3 the probability distribution  
$P(R_{\rm Abel})$  
of the abelian projection rate $R_{\rm Abel}$. 
Without gauge

\setlength{\miniwocolumn}{0.63\textwidth}

\begin{minipage}[t]{\miniwocolumn}

\begin{indention}{-0.6cm}
\vspace{-3.4cm}
fixing, the
average 
$\langle  R_{\rm Abel} \rangle$ is found to be about $2 \over 3$. 
In the MA gauge, the off-diagonal
 component of the SU(2) link variable 
is forced to be reduced, and
$R_{\rm Abel}$ approaches to unity;
one obtains   
$\langle  R_{\rm Abel} \rangle_{\rm MA} \simeq$  0.93
  on $16^4$ lattice
with $\beta = 2.4$.
Thus, we find  {\it microscopic abelian dominance} on the link variable.

\subsection{Analytical Study on Abelian Dominance for Confinement 
: Macroscopic Abelian Dominance}

\indent\indent In this section, 
we study the origin of abelian dominance on the confinement force,
considering the relation with  {\it  microscopic abelian dominance}
on the link variable\cite{suganuma1}.
In the MA gauge, the {\rm diagonal element} $\cos \theta_\mu(s)$ 
in $M_\mu(s)$ is maximized by the gauge transformation 
as large as possible;
$R_{\rm Abel}= 
0.93$ 
at $\beta=2.4$. Then, the {\it off-diagonal element 
$e^{i\chi_\mu(s)}\sin\theta_\mu(s)$ is forced to take a small value 
in the MA gauge,} and therefore the approximate treatment 
on the off-diagonal element would be allowed in the MA gauge.
Moreover, the {\it angle variable $\chi_\mu(s)$ is not 
constrained} by the MA gauge-fixing condition at all, 
and {\it tends to take a random value} besides the residual 
${\rm U(1)}_3$ gauge degrees 
of freedom.
Hence, $\chi_\mu(s)$ in the MA gauge can be regarded as a {\it random angle variable} 
in a good approximation. 

\end{indention}
\end{minipage}
\hspace{-0.5cm}
\setlength{\miniwocolumn}{0.33\textwidth}
\begin{minipage}[t]{\miniwocolumn}
\epsfxsize = 5 cm
\epsfbox{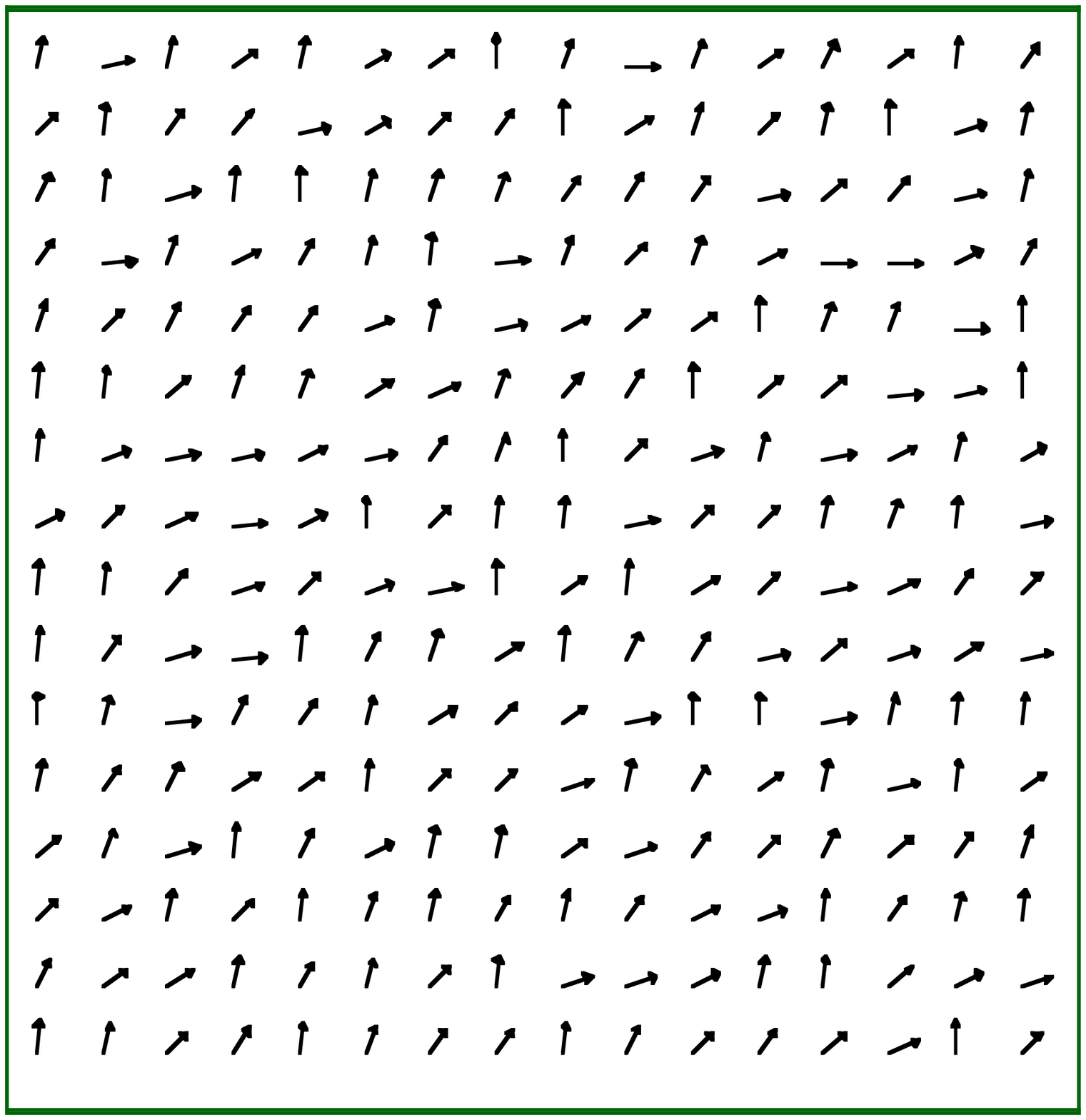}

\hspace{0.5cm}  { \footnotesize (a) No~Gauge~Fixing}

\vspace{.30cm}
\label{arrow5}
\epsfxsize = 5 cm
\epsfbox{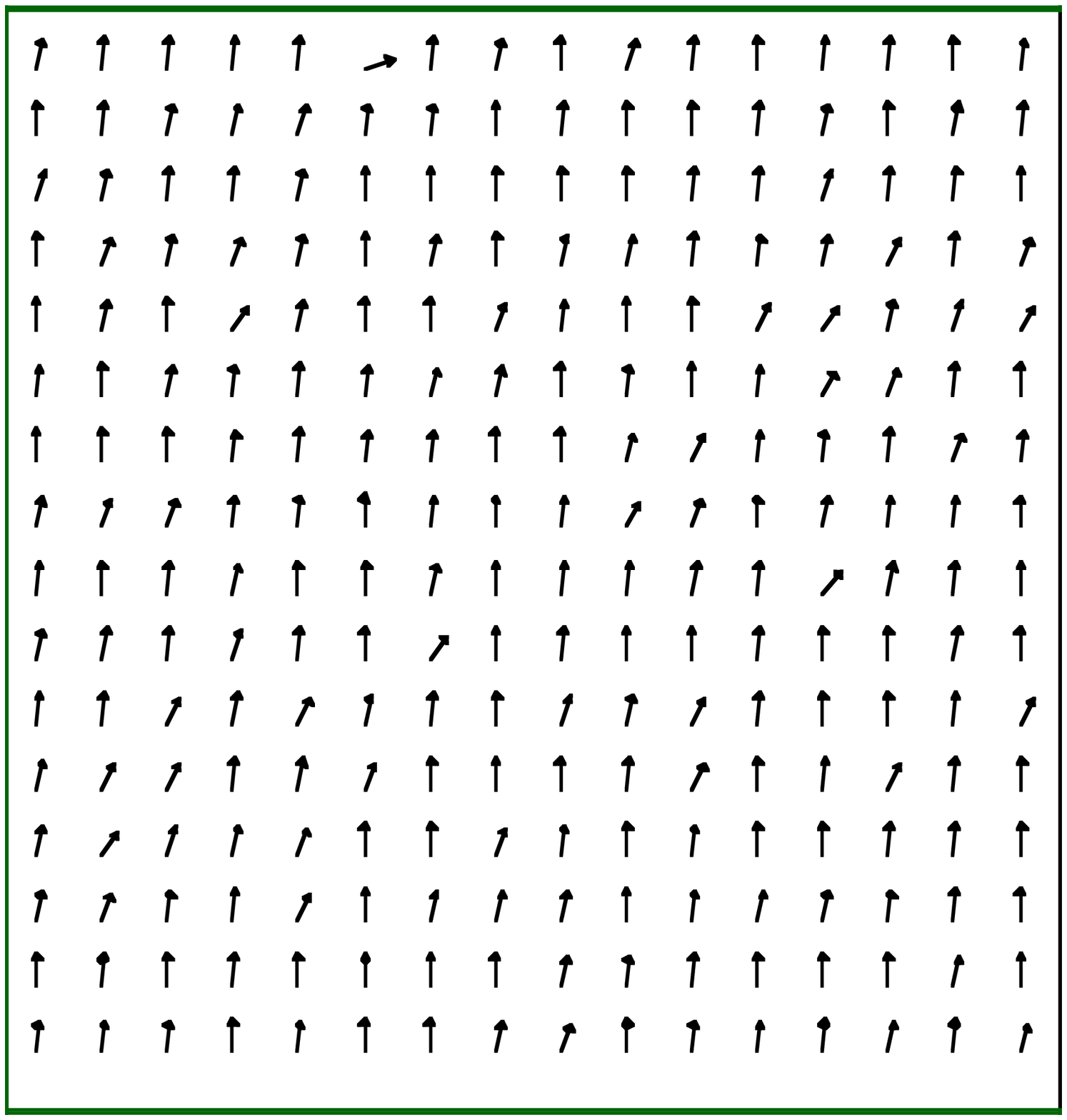}

\hspace{0.5cm} { \footnotesize (b) MA~Gauge~Fixing}

\vspace{0.1cm}
\begin{flushleft}
\begin{indention}{0.7cm}
{
 \footnotesize \noindent {\bf Fig.2} Local abelian projection rate 
$R_{\rm Abel}$ $=$ $\cos \theta$ 
at $\beta$=  2.4  on  
$16^4$ lattice. 
Arrows denote $(\sin \theta, \cos \theta) $ [$0 \le \theta 
\le \frac{\pi}{2}$].
}
\end{indention}
\end{flushleft}
\label{arrow6}
\vspace{0cm}
\end{minipage}

Let us consider the Wilson loop 
$\langle W_C[U_\mu(s)]\rangle \equiv 
\langle{\rm tr}\Pi_C U_\mu(s)\rangle$
in the MA gauge.
In calculating $\langle W_C[U_\mu(s)]\rangle$, 
the expectation value of $e^{i\chi_\mu(s)}$
in $M_\mu(s)$ vanishes as
$\langle e^{i\chi_\mu(s)}\rangle _{\rm MA}
\simeq \int_0^{2\pi} d\chi_\mu(s)\exp\{i\chi_\mu(s)\}=0$,
when $\chi_\mu(s)$ is assumed to be a {\it random angle variable.}
Then, the  off-diagonal factor 
$U_\mu(s)$ appearing 
in the Wilson loop  $W_C[U_\mu(s)]$ 
becomes a diagonal matrix,
$U_\mu(s)\equiv M_\mu(s)u_\mu(s)
\rightarrow 
\cos \theta_\mu(s) u_\mu(s),
$ in the MA gauge. 

Then, for the $I \times J$ rectangular $C$, the Wilson loop 
$W_C[U_\mu(s)]$ in the MA gauge is approximated as
\vspace{-.4cm}
{\small  
\begin{eqnarray}
\langle W_C[U_\mu(s)]\rangle &\equiv& 
\langle{\rm tr}  \prod_{i=1}^L U_{\mu_i}(s_i)\rangle
\simeq 
\langle  \prod_{i=1}^L \cos \theta_{\mu_i}(s_i) \cdot 
{\rm tr}   \prod_{j=1}^L u_{\mu_j}(s_j)\rangle_{\rm MA} \cr
\vspace{-.3cm}
& \simeq &
\langle\exp\{   \sum_{i=1}^L \ln (\cos \theta_{\mu_i}(s_i))\}\rangle_{\rm MA} 
\ \langle W_C[u_\mu(s)]\rangle_{\rm MA} \cr
\vspace{-.1cm}
&\simeq& 
\exp\{L \langle \ln (\cos \theta_\mu(s)) \rangle_{\rm MA} \} 
\ \langle W_C[u_\mu(s)]\rangle_{\rm MA},
\end{eqnarray}
\small }
\vspace{-.2cm}
\begin{indention}{-0.6cm}
\setlength{\miniwocolumn}{0.47\textwidth}

\begin{minipage}[t]{\miniwocolumn}
\epsfxsize = 5 cm
\epsfbox{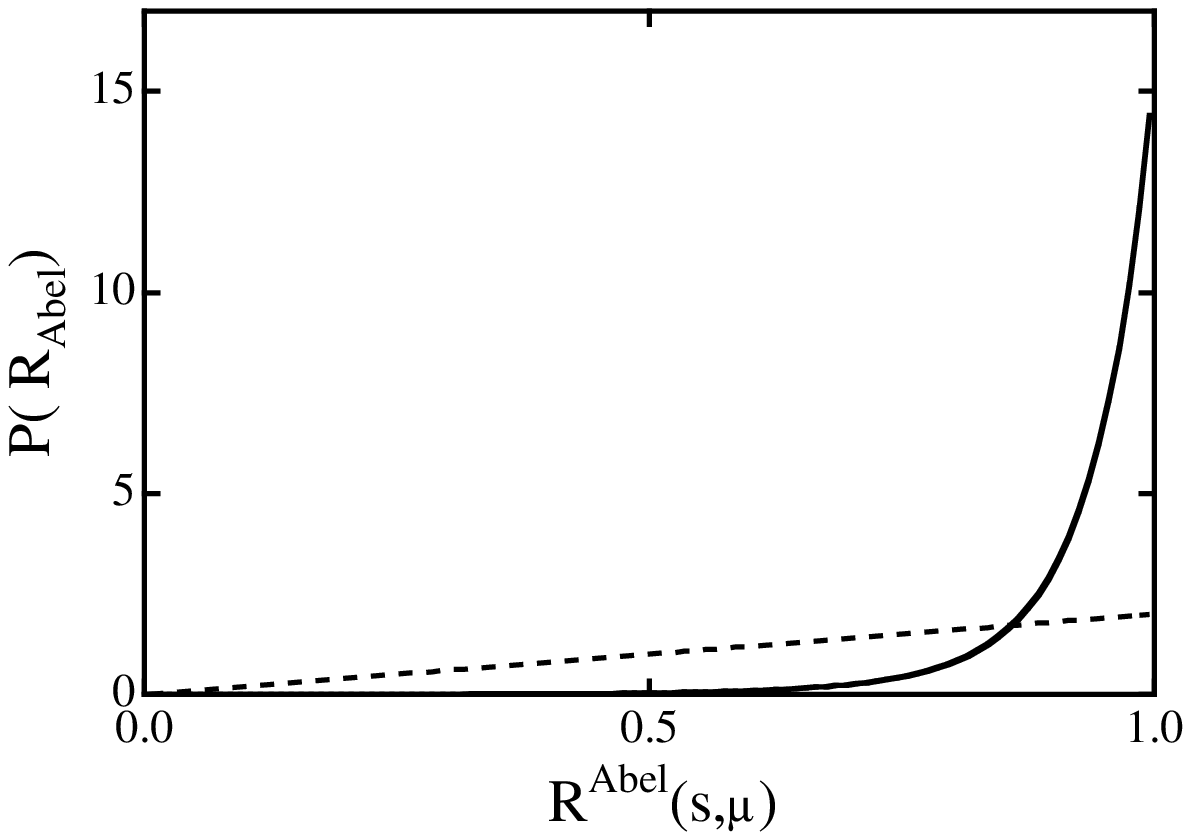}

{\small {\bf Fig.3}
 The probability distribution $P(R_{\rm Abel})$ 
of  abelian projection rate $R_{\rm Abel}(s,\mu) $
at $\beta =2.4$ on $16^4$ lattice from 40 gauge configurations
in the MA gauge (solid curve) and without gauge fixing (dashed curve).
}
\label{gfig3}
\vspace{0.3cm}
\end{minipage}
\hspace{0.3cm}
\begin{minipage}[t]{\miniwocolumn}
          \epsfxsize = 5.0cm
          \centerline{\epsfbox{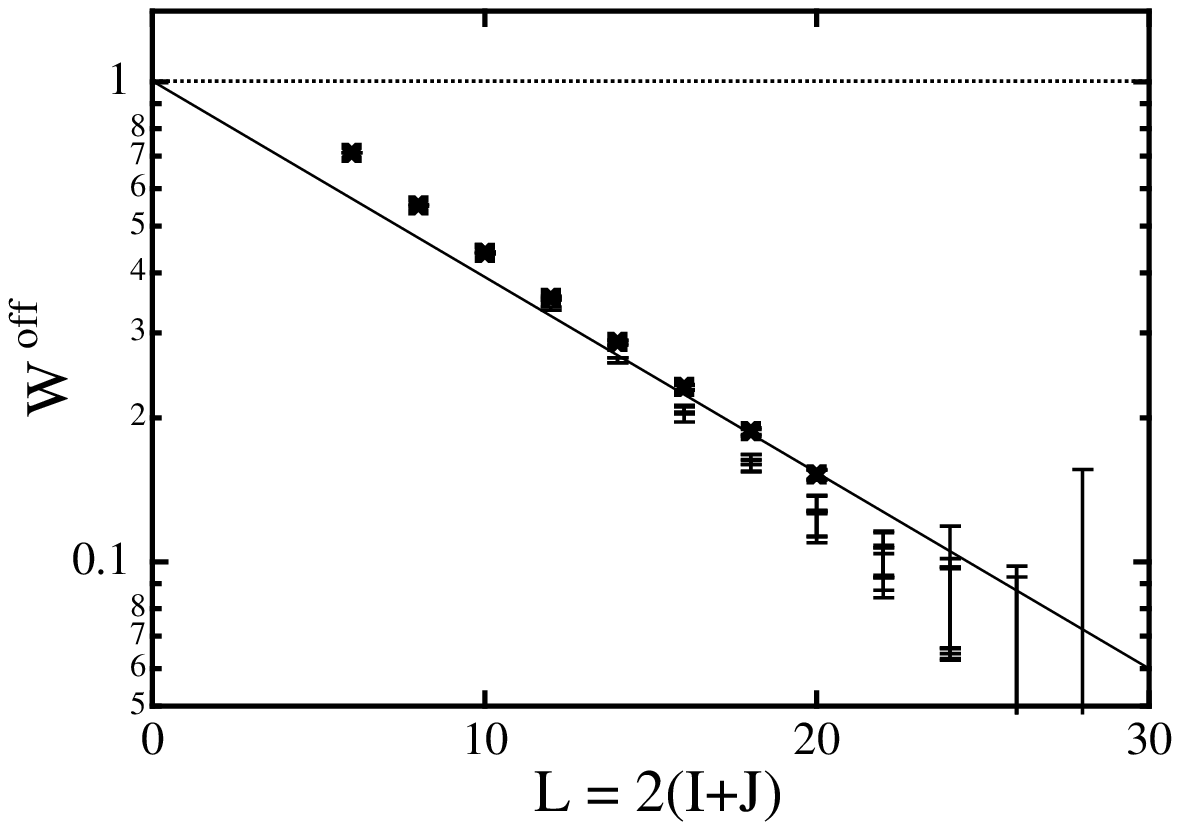}}

{\small {\bf Fig.4}
The comparison between 
the analytical estimation (straight line) and
the lattice data ($\times$)  
of $W_C^{\rm off}$
$\equiv$ 
$\langle W_C[U] \rangle/
\langle W_C[u]\rangle_{\rm MA}$ 
 with $I,J \ge 2$  
as the function of the perimeter $L \equiv 2(I+J)$ in the MA gauge
at $\beta=2.4$.
}
                                    \label{fig:off-w}
\end{minipage}
\end{indention}

\noindent where $L\equiv 2(I+J)$ denotes the perimeter length and 
$W_C[u_\mu(s)]\equiv {\rm tr}\Pi_{i=1}^L u_{\mu_i}(s_i)$ 
the abelian Wilson loop.
Here, we have replaced 
$\sum_{i=1}^L \ln \{\cos(\theta_{\mu_i}(s_i)\}$ 
by its average 
$L \langle \ln \{\cos \theta_\mu(s)\} \rangle_{\rm MA}$
 in a statistical sense.
In this way, we derive a simple estimation as 
$W_C^{\rm off}\equiv 
\langle W_C[U_\mu(s)]\rangle/\langle W_C[u_\mu(s)]\rangle_{\rm MA}
\simeq \exp\{L\langle \ln(\cos \theta_\mu(s))\rangle_{\rm MA}\}$
for the {\it contribution of the off-diagonal 
gluon element to the Wilson loop}.
From this analysis, the contribution of off-diagonal gluons 
to the Wilson loop is expected to obey the {\it perimeter law} 
in the MA gauge for large loops, when the statistical 
treatment would be accurate.

In the lattice QCD, we find that $W_C^{\rm off}$ seems to obey the 
{\it perimeter law} for the Wilson loop with $I,J \ge 2$ 
in the MA gauge, as shown in Fig.4. 
We find also that the lattice results on $W_C^{\rm off}$ 
as the function of $L$ is well reproduced 
by the above estimation with  {\it microscopic information} 
on the diagonal factor $\cos\theta_\mu(s)$ as 
$\langle \ln \{\cos_\mu(s)\} \rangle_{\rm MA}\simeq -0.082$ 
for $\beta=2.4$.

Thus, the off-diagonal contribution $W_C^{\rm off}$ 
to the Wilson loop obeys 
the perimeter law in the MA gauge, and therefore 
the SU(2) string-tension $\sigma_{\rm SU(2)}$ becomes
same as the abelian string-tension $\sigma_{\rm Abel}$,
\begin{eqnarray}
\sigma_{\rm SU(2)} & \equiv & 
-\lim_{I,J \rightarrow \infty}
{1 \over IJ}\ln \langle W_{I \times J}[U_\mu(s)]\rangle \nonumber \\
 & \simeq & -2 \langle \ln \{\cos\theta_\mu(s)\} \rangle_{\rm MA}
\lim_{I,J \rightarrow \infty} {I+J \over IJ}
+ \sigma_{\rm Abel}
\  \ \stackrel{{\small I,J \rightarrow \infty}}{\longrightarrow} \ \
\sigma_{\rm Abel},
\end{eqnarray}
when the finite size effect on $I$ and $J$ is removed. 
Thus, {\it abelian dominance for the string tension}, 
$\sigma_{\rm SU(2)}=\sigma_{\rm Abel}$, 
can be proved in the MA gauge by 
replacing the off-diagonal angle variable $\chi_\mu(s)$
as a random variable.

\section{Monopoles in QCD}

In this section, 
we derive the abelian gauge theory including monopoles
in terms of the gauge connection, 
and study monopole properties\cite{ichieA}. 

In the general system including singularities such as the Dirac string,
the field strength is defined as
$G_{\mu\nu}  \equiv \frac{1}{ie}
([{\hat D_\mu},{\hat D_\nu}]-[\hat \partial_\mu,\hat \partial_\nu])$
with the covariant derivative $\hat D_\mu \equiv \hat \partial_\mu + ie 
A_\mu$. 
By the general gauge transformation with the gauge function $\Omega $, 
the field strength $G_{\mu\nu}$ is transformed as
\begin{eqnarray}
G_{\mu\nu} \rightarrow G^\Omega_{\mu\nu}  =   
\Omega G_{\mu\nu} \Omega^{\dagger} 
& = & (\partial_\mu A^\Omega_\nu -\partial_\nu A^\Omega_\mu )
+ ie[A^\Omega_\mu, A^\Omega_\nu] +\frac{i}{e}
\Omega[\partial_\mu, \partial_\nu] \Omega^{\dagger}
\nonumber \\ 
& \equiv  & G^{\rm linear}_{\mu\nu} 
+  G^{\rm bilinear}_{\mu\nu} + G^{\rm sing}_{\mu\nu}.
\label{eq:field-strength}
\end{eqnarray}
The last term $G^{\rm sing}_{\mu\nu}$ remains only for the singular gauge transformation, 
and can provide the Dirac string. 
\begin{figure}[b]
\vspace{-1cm}
\epsfxsize = 8cm
\hspace{3cm}
\epsfbox{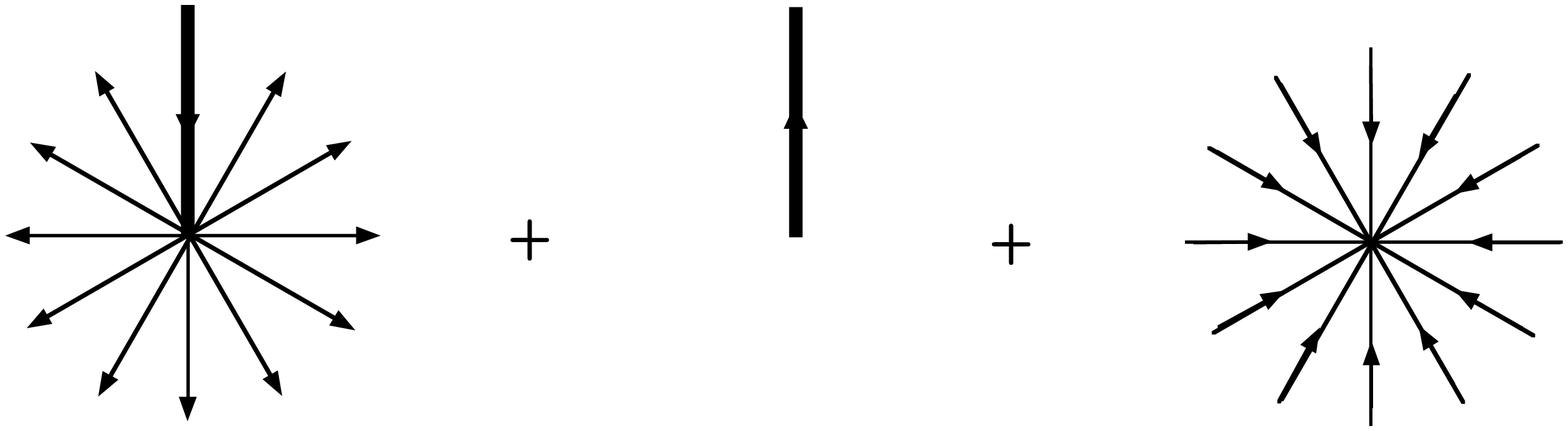}
\vspace{-3.5cm}
\begin{flushleft}
\begin{eqnarray}
 G^\Omega_{\mu\nu}  
 = 
 \hspace{4mm}
 \underbrace{
   \partial_\mu A^\Omega_\nu  - \partial_\nu A^\Omega_\mu
 \hspace{1mm}
   +   \hspace{1mm} \frac{i}{e}
 \Omega[\partial_\mu, \partial_\nu]  \Omega^{\dagger}
 \hspace{1mm} }   + \hspace{3mm} ie[A^\Omega_\mu, A^\Omega_\nu] 
\nonumber
 \end{eqnarray}
  
 \hspace{3.5cm} Abelian Projected QCD \\

 \hspace{1.5cm} {\small {\bf Fig.5} Appearance of monopoles in abelian projected QCD.
}
\vspace{0cm}
\end{flushleft}
\end{figure}

Here, in order to study appearance of the monopole, 
let us consider 
the SU(2) singular gauge-function
$\Omega = 
\left( {\matrix{
e^{i\varphi}{\rm cos}{\frac{\theta}{2}}  & {\rm sin}{\frac{\theta}{2}} \cr
-{\rm sin}{\frac{\theta}{2}} & e^{-i\varphi}{\rm cos}{\frac{\theta}{2}}}
} \right)
\label{eq:gauge-function}
$ 
with  polar angle 
$\theta$ and  azimuthal angle $\varphi$. 
Here, $ \Omega$ is multi-valued at the positive region of $z$-axis,
$\theta = 0$, 
and 
the last term $G^{\rm sing}_{\mu\nu}$ in Eq.(\ref{eq:field-strength})
provides the singular gauge configuration of the Dirac string
in the abelian sector,
as shown in Fig.5.
Accordingly, 
the linear term $G^{\rm linear}_{\mu\nu} $ 
and the second term $ G^{\rm bilinear}_{\mu\nu}$ 
in Eq.(\ref{eq:field-strength})
include 
{\it the monopole with the Dirac string} and the anti-monopole, respectively.
By the abelian projection $A_{\mu} \rightarrow {\cal A}_\mu \equiv 
A_{\mu}^3 T^3$, the bilinear term, which is originated
from the non-abelian nature, is dropped. The field-strength
in the abelian projected QCD is derived as
$ F_{\mu\nu} 
\equiv
\partial_\mu {\cal A}_\nu -\partial_\nu {\cal A}_\mu +\frac{i}{e}
 \Omega[\partial_\mu, \partial_\nu]  \Omega^{\dagger}$, which
breaks the abelian Bianchi identity.
Thus, by the singular gauge transformation as $\Omega$,
 the monopole appears in the abelian 
sector of QCD.

Next, we consider  monopole properties in terms of the action.
In the static frame of the monopole 
with the magnetic charge $g$, 
a spherical `magnetic field' is created around the monopole 
in the abelian sector of QCD as 
${\bf H}(r) =  \frac{g}{4\pi r^3} {\bf r}$
with ${\bf H}_i\equiv \epsilon _{ijk} \partial_j A^3_k$. 
Then, the monopole inevitably accompanies 
a large fluctuation of the abelian gluon component $A^3_\mu$ around it. 
For the abelian part $S_{\rm Abel}\equiv -\frac14 \int d^4x
(\partial_\mu A_\nu^3-\partial_\nu A_\mu^3)^2$ of the QCD action, 
the electro-magnetic energy
created around the monopole 
is estimated as 
$E(a) = \int_a^\infty d^3 x \frac12 {\bf H}(r)^2 = 
\frac{g^2}{ 8 \pi a},$
where $a$ is an ultraviolet cutoff like a lattice mesh.
As the ``mesh'' $a$ goes to 0, 
the monopole provides a large fluctuation of 
$S_{\rm Abel}$, 
and hence the monopole seems difficult to appear if the abelian 
gauge theory is controlled by $S_{\rm Abel}$.
This is the reason why QED does not have the point-like Dirac monopole.
Then, why can the monopole appear in the abelian sector of QCD ?
To answer it, let us consider the division of the total QCD action 
$S_{\rm QCD}$ into the abelian part $S_{\rm Abel}$ and 
the remaining part $S_{\rm off}\equiv S_{\rm QCD}-S_{\rm Abel}$, which 
is contribution from the off-diagonal gluon component. 
Unlike $S_{\rm QCD}$ and $S_{\rm Abel}$, 
$S_{\rm off}$ is not positive definite and can take a negative value
in the Euclidean metric.
Then, around the monopole, 
the abelian action $S_{\rm Abel}$ should be 
partially canceled by the remaining  off-diagonal contribution $S_{\rm off}$
such that the total QCD action $S_{\rm QCD}$ around the monopole 
does not become extremely large. 
Thus, we expect large off-diagonal gluon components 
around the monopole for  cancellation with the
large field fluctuation of the abelian part.
Based on this analytical consideration, we study
action densities around monopoles in the MA gauge using the lattice QCD.

On the SU(2) lattice, we measure the action densities 
$\bar S_{\rm SU(2)}$, $\bar S_{\rm Abel}$ and $\bar S_{\rm off}$,
which are the SU(2), the abelian and the off-diagonal parts, respectively. 
We show 
the probability distributions 
of 
$\bar S_{\rm SU(2)}$, $\bar S_{\rm Abel}$ and $\bar S_{\rm off}$ in Fig.6.
(To be exact, $\bar S$ is the averaged value over the neighboring links 
around a dual link\cite{suganuma1}.)
 For the total distribution on the whole lattice,  most 
$\bar S_{\rm off}$ are positive, and both $\bar S_{\rm Abel}$ and 
$\bar S_{\rm off}$ tend to take smaller values than $\bar S_{\rm SU(2)}$
owing to 
\begin{indention}{-0.6cm}
\setlength{\miniwocolumn}{0.47\textwidth}
\vspace{0.3cm}
\begin{minipage}[b]{\miniwocolumn}
\epsfxsize= 50mm
\epsfbox{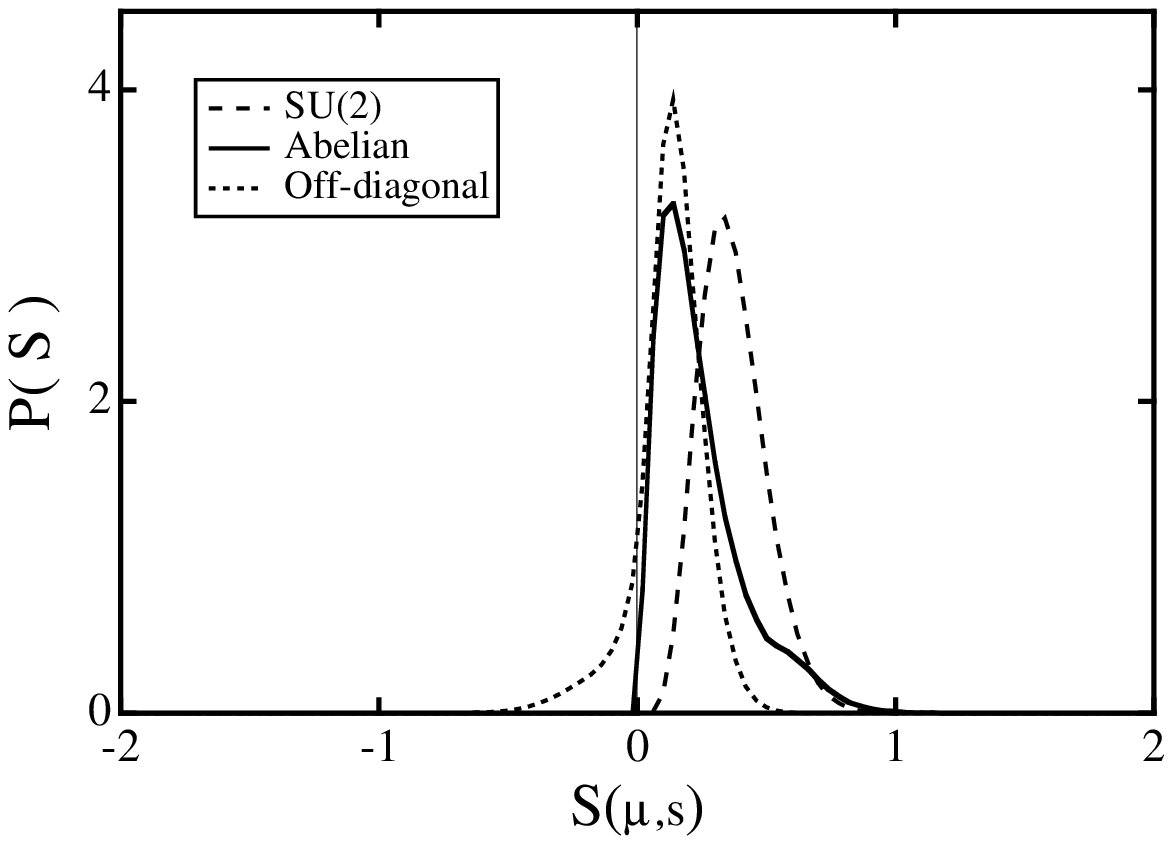}

\noindent
\baselineskip 9pt
 \footnotesize
{\bf Fig.6(a)} Probability distributions $P(\bar S)$ of the 
 $\bar S_{\rm SU(2)}$(dashed curve), 
$\bar S_{\rm Abel}$ (solid curve) and 
$\bar S_{\rm off}$
(dotted curve) in the MA gauge at $\beta =2.4$ on $16^4$ lattice.
\end{minipage}
\hspace{0.3cm}
\begin{minipage}[b]{\miniwocolumn}
\epsfxsize=50mm
\epsfbox{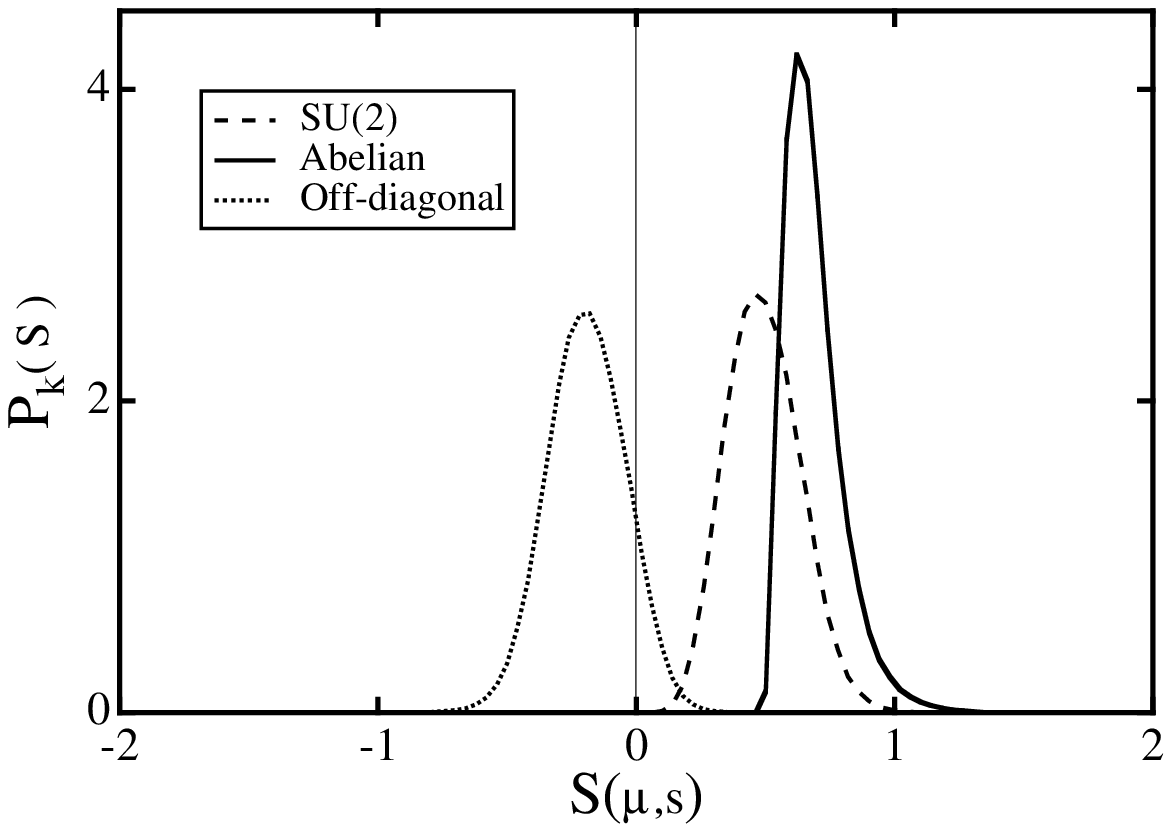}

\noindent
\baselineskip 9pt
{ \footnotesize
{\bf Fig.6(b)} Probability distribution $P_k(\bar S)$ 
{\bf around the monopole} for $\bar S_{\rm SU(2)}$, $\bar S_{\rm Abel}$ 
and $\bar S_{\rm off}$
in the MA gauge at $\beta =2.4$ on $16^4$ lattice.
 }
\end{minipage}
\end{indention}
\noindent
$\bar S_{\rm SU(2)} = \bar S_{\rm Abel} +  \bar S_{\rm off}$. 
Around the monopole, however, the off-diagonal part $\bar S_{\rm off}$ of 
the action density tends to take a large negative value, and  
$ \bar S_{\rm off}$ strongly cancels with the large abelian action 
density  $\bar S_{\rm Abel}$ so as to keep the total SU(2) action small.
Thus, monopoles can appear in the abelian sector in QCD without large cost 
of the QCD action due to the strong cancellation between the abelian 
action $ S_{\rm Abel}$ and the off-diagonal part $ S_{\rm off}$ 
of the action.

In conclusion, in the small distance, 
all SU(2) components contribute to the short-distance 
physics. In the dual-superconductor picture, the off-diagonal components
are necessary to create the monopole in the abelian sector.
In the large distance, however, 
the off-diagonal components are not necessary for the infrared physics, and
only abelian gauge fields with monopoles describe the QCD vacuum.

\vspace{0.5cm}

\setlength{\miniwocolumn}{0.63\textwidth}
\begin{minipage}[h]{\miniwocolumn}

\begin{indention}{-0.6cm}

\section{Monopole Condensation}

In the dual Higgs theory, the monopole is assumed to be condensed, 
which has been suggested by the formation of 
the
global network of the 
monopole current in the lattice QCD.
In this section, we consider monopole condensation 
in the QCD-vacuum at the infrared scale.
However, QCD is described by the gluon field not by the monopole 
current, and therefore it is difficult to clarify  monopole 
condensation only with the lattice QCD simulation.
To this end, we generate the monopole-current system on the lattice 
using a simple monopole-current action, and study 
monopole condensation and 
the role of the 
monopole to color confinement\cite{ichie2}.

\indent\indent\indent \hspace{0.5cm} In general, monopole-current action
$S[k_\mu(s)]$ includes the nonlocal
Coulomb interaction as $S_C=\int d^4x d^4y k_\mu(x) D(x-y) k_\mu(y)$
with the Coulomb propagator $D(x)$. In the dual Higgs phase, however,
the effective interaction between monopole currents would be
short-range due to the {\it screening effect} by the dual Higgs mechanism
\cite{suganuma1,ezawa,atanaka}
similar to the Debye screening\cite{stack}. 
Then, the infrared partition functional of the monopole current
$k_\mu(x) \equiv k_\mu^3(x) \cdot \frac{\tau^3}{2}$ would be written as
\end{indention}
\end{minipage}
\hspace{-0.5cm}
\setlength{\miniwocolumn}{0.4\textwidth}
\begin{minipage}[h]{\miniwocolumn}
\epsfxsize = 5 cm
\epsfbox{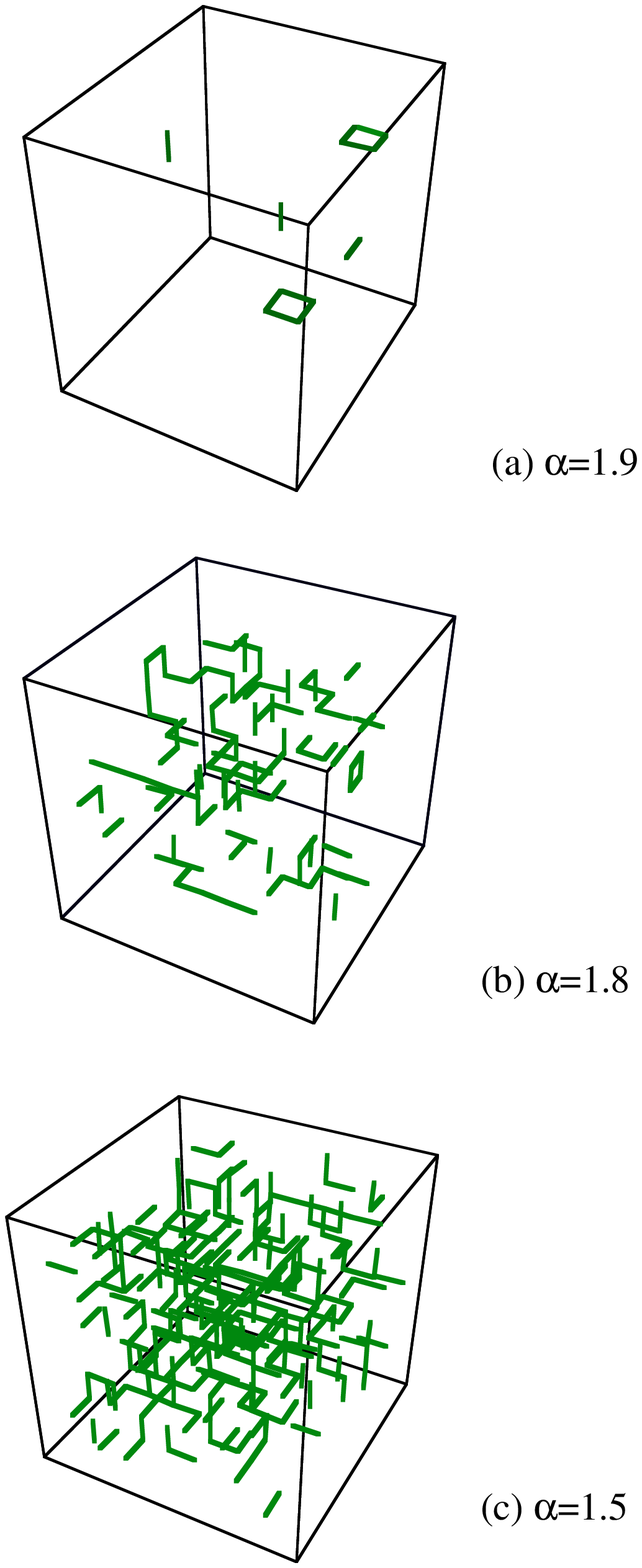}
\begin{flushleft}
\begin{indention}{0.7cm}
\baselineskip 9pt
{\small  
{\bf Fig.7}  Monopole current  \\ in the 
monopole-current system in ${\bf R}^3$ at a fixed time.
}
\label{fig:mcfig1}
\end{indention}
\end{flushleft}
\vspace{0cm}
\end{minipage}
\begin{equation}
Z = \int {\it D}k_\mu {\rm exp} \{ -\alpha \int_{a} d^4x 
{\rm tr}  k^2_\mu (x) \}
\delta(\partial_\mu k_\mu),
\label{eq:part}
\end{equation}
where $\alpha$ is the energy per unit length of the monopole current.
Here, 
$a$ is an ultraviolet cutoff larger than the screening length.
We put the system on the 4-dimensional lattice with the lattice spacing 
$a$. 
Fig.7 shows behavior of the monopole current in the monopole
current system for $\alpha =1.5,1.8,1.9$.
When energy $\alpha$ becomes smaller, the monopole current is generated 
more easily.
As $\alpha=1.5$, the monopole current covers the whole lattice and 
creates the 
global network, which is similar to the monopole current in the QCD vacuum.

Here, let us consider 
this behavior of monopole currents in the analytical way.
In the abelian gauge of QCD, the charge of the monopole is expected 
to be unity\cite{suganuma1}. 
The partition function can be approximated 
as the single monopole-loop ensemble with the length $L$, 
$Z=\sum_{L}\rho(L)e^{-\alpha L}$,
where $L$ and $\rho(L)$ are length of the monopole loop and its 
configuration number, respectively. 
The monopole current with the length $L$ is regarded as the $L$ step
self-avoiding random walk, where $2d-1=7$ direction is possible 
in each step.
Therefore, $\rho(L)$ is roughly estimated as $(2d-1)^L=7^L$,
and 
the expectation value of the monopole-current length is found to be 
\begin{equation}  
\langle L \rangle =\frac{1}{Z}\sum_{L}\rho(L) L e^{-\alpha L} =
  \left\{ \begin{array}{ll}
  \{ \alpha-\ln(2d-1) \}^{-1}  & \mbox{if $\alpha > \ln(2d-1)$} \\
  \infty & \mbox{if $\alpha < \ln(2d-1)$}.
       \end{array}
  \right. 
\end{equation}
When  energy $\alpha$ is larger 
than ``entropy'' $\ln(2d-1)$, the monopole-loop length is finite. 
However, when $\alpha$ is smaller than entropy, 
the monopole-loop length 
becomes infinite, which corresponds to monopole condensation 
in the current representation\cite{ezawa}.
Here, the critical value on 
monopole condensation  is $\alpha_{c} \simeq {\rm ln}(2d-1) \simeq {\rm 
ln}7 \simeq 1.945$,
which corresponds to  ``entropy'' of the self-avoiding random 
walk.
Such a transition is quite similar to the Kosterlitz-Thouless
transition in 2-dimensional superconductors, 
where vortex condensation plays an important role to the transition.

Next, we study how these monopole currents 
contribute to color confinement properties\cite{ichie2}.
Quark confinement is characterized by the linear inter-quark potential,
which can be obtained from  area-law behavior of the Wilson 
loop,
$\langle W \rangle 
= \langle P {\rm exp}({ie\oint A_\mu dx_\mu}) \rangle$.
The expectation value of the Wilson loop $\langle W \rangle$ is shown in
Fig.8.
The Wilson loop exhibits the area law and the linear confinement 
potential:
${\rm ln} \langle W \rangle$ decreases linearly with the 
quark loop area $I \times J$.
Quantitatively, the string tension is measured by the Creutz ratio,
and we show in Fig.9 $\chi(3,3)$ as a typical example.
For the monopole condensed phase as $\alpha < \alpha_c$, the string 
tension gets a finite value, while it vanishes for the non-condensed 
phase of monopoles as $\alpha \ge \alpha_c$.
Thus, the confinement phase directly corresponds to the monopole 
condensed phase, and therefore monopole condensation is considered 
as essence of the confinement mechanism.

\begin{indention}{-0.6cm}
\setlength{\miniwocolumn}{0.47\textwidth}
\vspace{0.3cm}
\begin{minipage}[t]{\miniwocolumn}
\epsfxsize= 50mm
\epsfbox{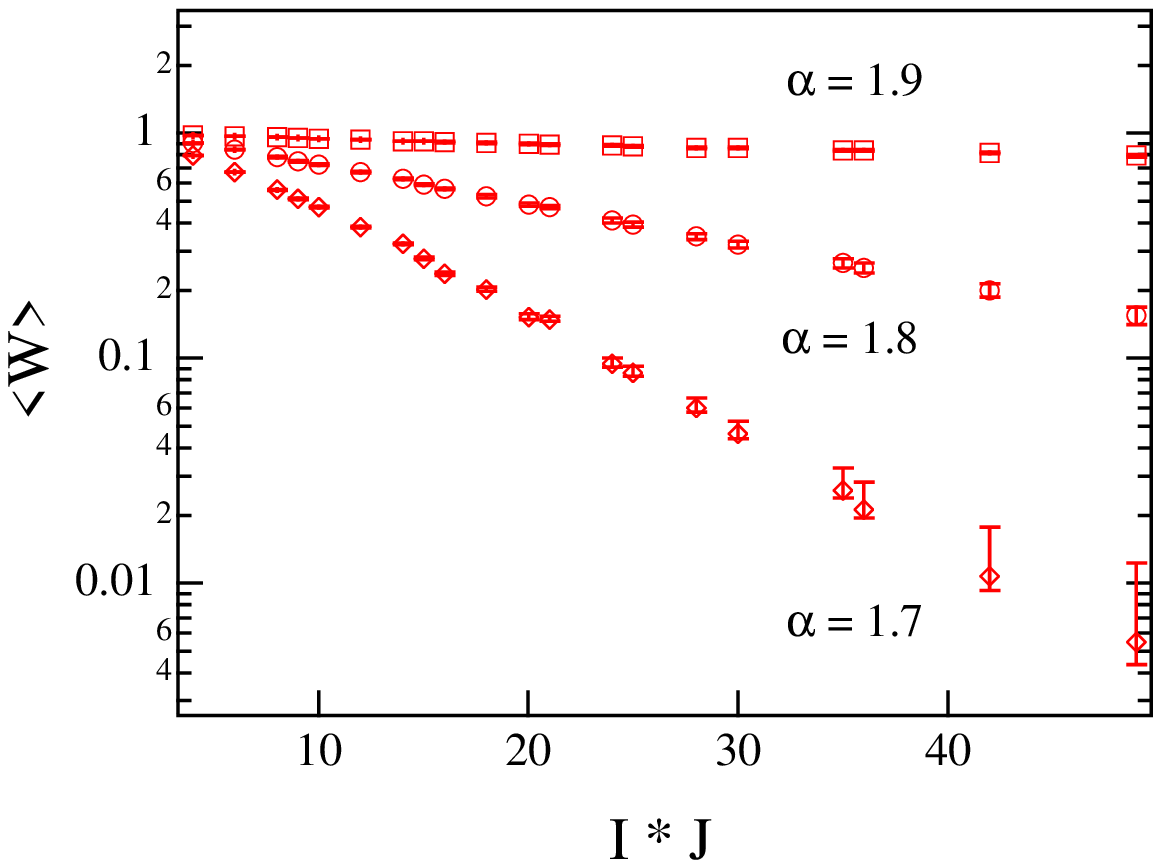}
\vspace{0cm}

\noindent
\baselineskip 9pt
{\footnotesize {\bf Fig.8}
The Wilson loop
$\langle W (I \times J) \rangle$ 
in the monopole-current system.} 
\label{fig:mcfig5}
\vspace{0cm}
\end{minipage}
\hspace{0.3cm}
\begin{minipage}[t]{\miniwocolumn}
\epsfxsize=50mm
\epsfbox{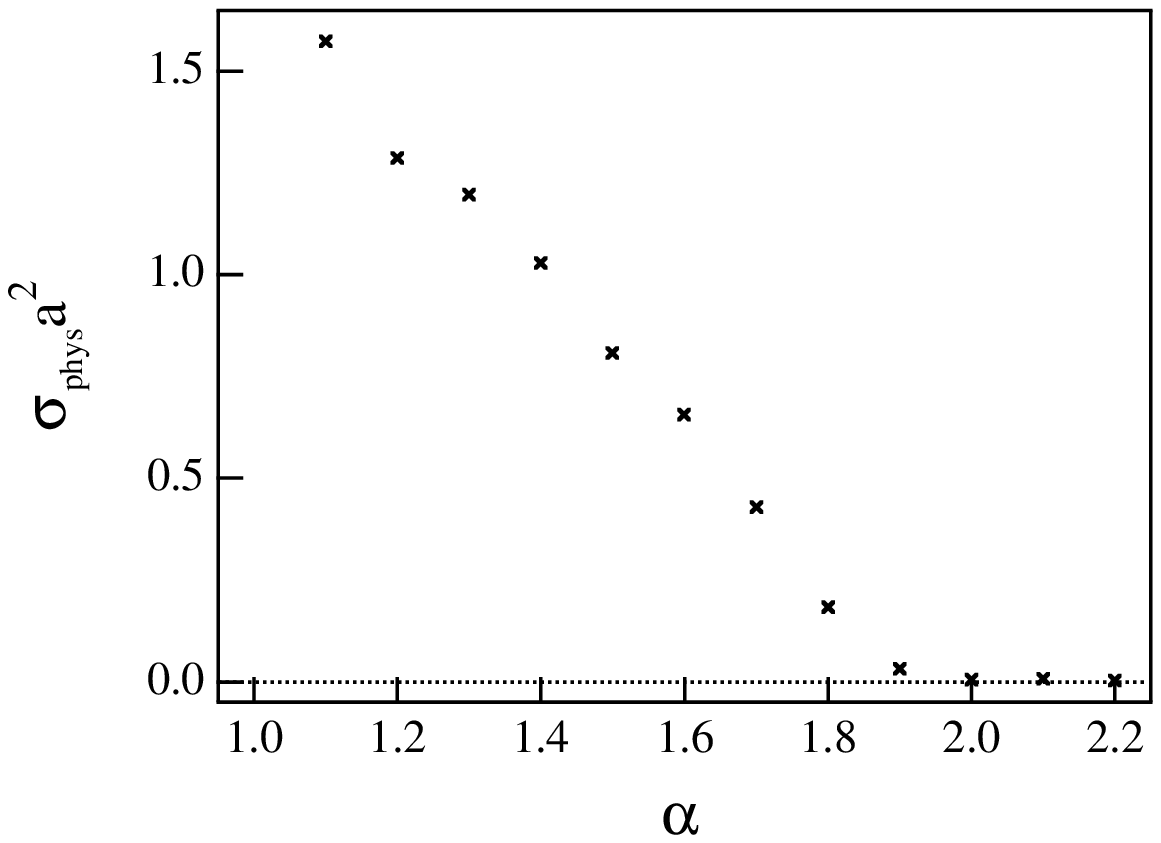}

\noindent
\baselineskip 9pt
{\footnotesize {\bf Fig.9}
 The Creutz ratio as the function of $\alpha$ in the multi-monopole system.
}
\label{fig:mcfig6}
\vspace{0.5cm}
\end{minipage}
\end{indention}

We compare the lattice QCD with the monopole-current 
system in terms of monopole condensation and confinement properties.
The lattice QCD simulation shows that the QCD vacuum in the MA gauge 
holds the global network of  the monopole-current. 
Considering the similarity on the monopole clustering, the QCD vacuum 
can be regarded as the monopole condensed phase with $\alpha < \alpha_{c}$
in the monopole-current system, as shown in Fig.7(c). 
Such an identification of the QCD vacuum with the monopole condensed 
phase is also suggested 
in terms of the confinement properties, because the confinement phase 
corresponds to the monopole condensed phase as shown in 
Fig.9\cite{ichie2}.
Thus, monopole condensation is considered to occur 
at the infrared scale in the QCD vacuum 
and the confinement would be caused by monopole condensation there.

One of authors (H.I.) is supported by 
Research Fellowships of the Japan Society for the 
Promotion of Science for Young Scientists.

\end{document}